\documentclass[aps,twocolumn,showpacs,preprintnumbers,prb]{revtex4}

\usepackage{graphicx}
\usepackage{dcolumn}
\usepackage{bm}

\begin{document}


\title{Large crystal local-field effects in the dynamical structure
factor of
rutile TiO$_2$}
\author{I.~G.~Gurtubay$^{1}$, Wei Ku$^2$, J.~M.~Pitarke$^{1,3}$, 
A.~G.~Eguiluz$^{4,5}$, B.~C.~Larson$^5$, J.~Tischler$^5$, and
P.~Zschack$^6$}
\affiliation{
$^1$Materia Kondentsatuaren Fisika Saila, Zientzi Fakultatea,
Euskal Herriko Unibertsitatea,\\
644 Posta kutxatila, E-48080 Bilbo, Basque Country, Spain\\
$^2$Department of Physics, Brookhaven National Laboratory,
Bldg 510, Upton, NY 11973-5000\\
$^3$Donostia International Physics Center (DIPC) and Unidad F\'\i sica
Materiales
CSIC-UPV/EHU,\\ 
Manuel de Lardizabal Pasealekua, E-20018 Donostia, Basque Country,
Spain\\
$^4$Department of Physics and Astronomy, The University of Tennessee,
Knoxville, Tennessee 37996-1200\\
$^5$ Condensed Matter Sciences Division, Oak Ridge National
Laboratory, 
Oak Ridge, Tennessee 37831-6030\\
$^6$ Frederick Seitz Materials Research Laboratory, University of
Illinois,
Urbana-Champaign,
Illinois 61801}

\date{\today}

\begin{abstract}
We present {\it ab initio} time-dependent-density-functional
calculations and
non-resonant inelastic x-ray scattering measurements of the
dynamical
structure factor of rutile TiO$_2$. Our calculations are in good
agreement with
experiment and prove the presence of large crystal local-field
effects below the Ti
M-edge, which yield a sharp loss peak at 14~eV whose intensity
features a remarkable
non-monotonic dependence on the wave vector. These effects, which
impact the
excitation spectra in the oxide more dramatically than in transition
metals,
provide a signature of the underlying electronic structure.
\end{abstract}

\pacs{71.15.Mb, 71.45.Gm, 78.70.Ck}

\maketitle

Titanium dioxide (TiO$_2$) has been studied extensively for its
remarkable
electric, magnetic, catalytic, and electrochemical properties. Based
on these
properties, TiO$_2$ has been used in a wide variety of technological
applications, such as dielectric material for integrated electronics
and
photocatalyst for the decomposition of organic compounds.\cite{asahi} 

In the last decade, many investigations
have been 
carried out focusing on the structural, electronic, and optical
properties
of this transition-metal oxide. From the experimental point of view it
has been
studied by various techniques, such as ultraviolet
photoemission,\cite{tio2ups,tio2ups2} x-ray emission,\cite{tio2xes}
x-ray
photoemission, \cite{tio2xps,tio2xps2} Auger-electron,\cite{tio2aes}
and electron-energy-loss spectroscopies.
\cite{tio2eels,tio2eels2,tio2launay} Theoretical investigations of
this wide-band-gap semiconductor include {\it ab initio} calculations
of
its structural, electronic, and optical
properties.\cite{tio2glass2,tio2mo}

Recently, Launay {\it et al.}\cite{tio2launay} have shown evidence of
a rutile-phase
characteristic peak in the low-energy loss spectra of TiO$_2$, and
Vast {\it et
al.}\cite{tio2reining} have investigated the impact of crystal
local-field effects (LFE) on
the electron energy-loss spectra of rutile TiO$_2$. These authors used
electron energy
loss spectroscopy (EELS) to study the low wave vector behaviour of the
energy-loss
function of this material. They combined their experimental spectra
with {\it ab initio}
calculations, and concluded that local-field effects are only
relevant at energies
above 40~eV, where excitations from the Ti $3p$ semicore levels
occur.       

In this paper, we report {\it ab initio}
time-dependent-density-functional
(TDDFT) calculations of the dynamical structure factor of rutile
TiO$_2$ at large
momentum transfers. We also report non-resonant inelastic x-ray
scattering (IXS)
measurements, which are suited to study the short-wavelength behaviour
of this
quantity, and we observe large crystal local-field effects at low
energies. These effects are absent at the small momentum transfers
accesible by EELS;\cite{tio2reining} however, we find that at larger
momentum transfers, where inhomogeneities due to localized $d$-states
can be sampled, crystal local-field effects yield a  sharp
non-monotonic loss peak at 14~eV.

Within the first Born approximation, the  inelastic scattering
cross-section
for x-rays to transfer momentum $\hbar({\bf q}+{\bf G})$ and energy
$\hbar\omega$ to a periodic solid is characterized by the
dynamical structure factor of the solid:
\begin{equation}\label{sqw}
{d^2\sigma\over d\Omega d\omega}=\left({e^2\over
m_ec^2}\right)^2\,({\bf
e}_i\cdot{\bf
e}_f)^2\,\left({\omega_f\over\omega_i}\right)\,\hbar\,S({\bf
q}+{\bf G},\omega),
\end{equation}
where (${\bf e}_i$,${\bf e}_f$) and ($\omega_i$,$\omega_f$) refer to
the
polarization vector and frequency of the incident and scattered
photon,
respectively, the wave vector ${\bf q}$ is in the first Brillouin zone
(BZ),
and ${\bf G}$ is a vector of the reciprocal lattice. $S({\bf q}+{\bf
G},\omega)$ represents the dynamical structure factor\cite{pines}
\begin{equation}
S({\bf q}+{\bf G},\omega)=-2\,\Omega\,{\rm Im}\chi_{{\bf G},{\bf
G}}({\bf
q},\omega).
\end{equation}
Here, $\chi_{{\bf G},{\bf G}'}({\bf q},\omega)$ is the
density-response
function of the solid, and $\Omega$ is the normalization volume.

In the framework of TDDFT,\cite{Petersilka-96} 
the {\it exact} density-response matrix of a periodic solid of
electron density
$n_0({\bf r})$ can be written as
\begin{eqnarray}\label{eq:XGG}
&&\chi_{{\bf G},{\bf
G}'}({\bf q},\omega)=\chi^S_{{\bf G},{\bf G}'}({\bf q},\omega)+
\sum_{{\bf G}_1,{\bf G}_2}\,\chi^S_{{\bf G},{\bf G}_1}({\bf
q},\omega)\cr\cr&&\times\left\{v_{{\bf G}_1}({{\bf q}})+f_{{\bf
G}_1,{\bf G}_2}^{\rm XC}[n_0]({\bf q},\omega)\right\}\chi_{{\bf
G}_2,{\bf
G}'}({\bf q},\omega).
\end{eqnarray}
Here, $f_{{\bf G},{\bf G}'}^{\rm XC}[n]({\bf q},\omega)$ are the
Fourier
coefficients of the functional derivative of the time-dependent
exchange-correlation (XC) potential of TDDFT, and
$\chi^S_{{\bf G},{\bf G}'}({\bf q},\omega)$ represents the
density-response
matrix of noninteracting Kohn-Sham (KS) electrons,\cite{eguiluz} which
is
obtained from the knowledge of the eigenfunctions and eigenvalues of
the
single-particle Kohn-Sham equation of density-functional theory
(DFT).\cite{kohn64,kohn65} We solve this equation self-consistently in
the
local-density approximation (LDA), with use of the Perdew-Zunger
parametrization\cite{perdew} of the Ceperley-Alder XC energy of a
uniform
electron gas.\cite{ca} In the random-phase approximation (RPA), the XC
kernel
$f_{{\bf G},{\bf G}'}^{\rm XC}[n]({\bf q},\omega)$ entering
Eq.~(\ref{eq:XGG})
is taken to be zero; in the so-called adiabatic LDA (ALDA), it is
approximated by
an adiabatic local kernel of the form
\begin{equation}\label{fxcgg}
f_{{\bf G},{\bf G}'}^{\rm XC}({\bf
q},\omega)=\int d{\bf r}\,{\rm e}^{-i({\bf G}-{\bf G}')\cdot{\bf
r}}\,\left.{dV_{\rm XC}(n)\over
dn}\right|_{n=n_0({\bf r})},
\end{equation}
where $V_{\rm XC}(n)$ is the XC potential of a homogeneous electron
gas of
density $n$.

In the calculations presented below, we first expand the LDA
single-particle
Bloch states in a linearized augmented plane wave (LAPW)
basis,\cite{wien} by
dividing the unit cell into non-overlapping atomic spheres and the
interstitial region between them. Inside the atomic spheres, the LAPW
wave
functions are expanded in spherical harmonics with $l_{\rm max}=10$.
The Ti
3$s$ and 3$p$ states and the O 2$p$ states are treated as semi-core
states,
and the Kohn-Sham equation is solved with a cutoff parameter
$R_{MT}K_{max}=8$
and an energy cutoff of 7.5 Ry. The Kohn-Sham density-response matrix
has been
computed as in Ref.~\onlinecite{Ku02}, with a damping parameter  of
0.2 eV.

Non-resonant inelastic x-ray scattering measurements of the dynamical
structure factor
of rutile TiO$_2$ were obtained using the UNICAT undulator Beam Line
on the Advanced Photon Source at Argonne National Laboratory. 
The measurements were made in reflection geometry, with 7.6 keV
x-rays,
a range of
wave vectors from 0.5 to 3.98 ${\rm\AA}^{-1}$ in the [001] direction,
and an energy
resolution of $1.1\,{\rm eV}$. The surprising result of these
measurements is the non-monotonic wave-vector dependence of the
intensity of a sharp loss peak at $14\,{\rm eV}$, which is absent in
existing optical ($q$=0) and EELS measurements on
this material and whose physics is qualitatively different from the
well-known
collective excitations in simple metals. Our {\it ab initio}
calculations indicate that this feature is the result of large crystal
local-field effects near the plasma frequency. 

In order to investigate the impact of crystal local-field effects on
the
dynamical structure factor of rutile TiO$_2$, we have first neglected
these effects
by considering only the diagonal elements of the Kohn-Sham
density-response
matrix entering Eq.~(\ref{eq:XGG}) (diagonal calculation), and we have
then solved this
matrix equation with a given number of ${\bf G}$
vectors (full calculation).\cite{LFEfoot}
Well converged results have
been obtained with the matrix size ranging from $63\times 63$ to
$113\times
113$, depending on the momentum transfer and the energy range
considered,
 and using 30 points in the irreducible BZ (IBZ). 

The energy-loss function ${\rm Im}\left[-\epsilon_{{\bf 0},{\bf
0}}^{-1}({\bf q},\omega)\right]$\cite{epsilonfoot}
 of rutile TiO$_2$ for vanishing momentum transfer
(long wavelengths) has been reported by Mo and Ching\cite{tio2mo} with
no
inclusion of crystal local-field effects and more recently by Vast
{\it et
al.}\cite{tio2reining} with full inclusion of these effects. At small
wave vectors, two principal structures can be indentified in the
energy-loss
function at energies below the titanium M-edge: a plasmon
peak at 12~eV, and a broad collective excitation at about 25 eV,
stemming
from the
building up of collective modes of the strongly hybridized Ti $3d$ and
O $2p$
bands. Vast {\it et al.}\cite{tio2reining} also reported
loss-function calculations of rutile TiO$_2$ for $|{\bf q}|\approx
0.4\,{\rm\AA}^{-1}$, showing that at these small wave vectors crystal
local-field effects have little impact at energies below the titanium
M-edge but drastically reduce the peak heights above the Ti M-edge.

\begin{figure*}
\includegraphics*[width=0.93\linewidth]{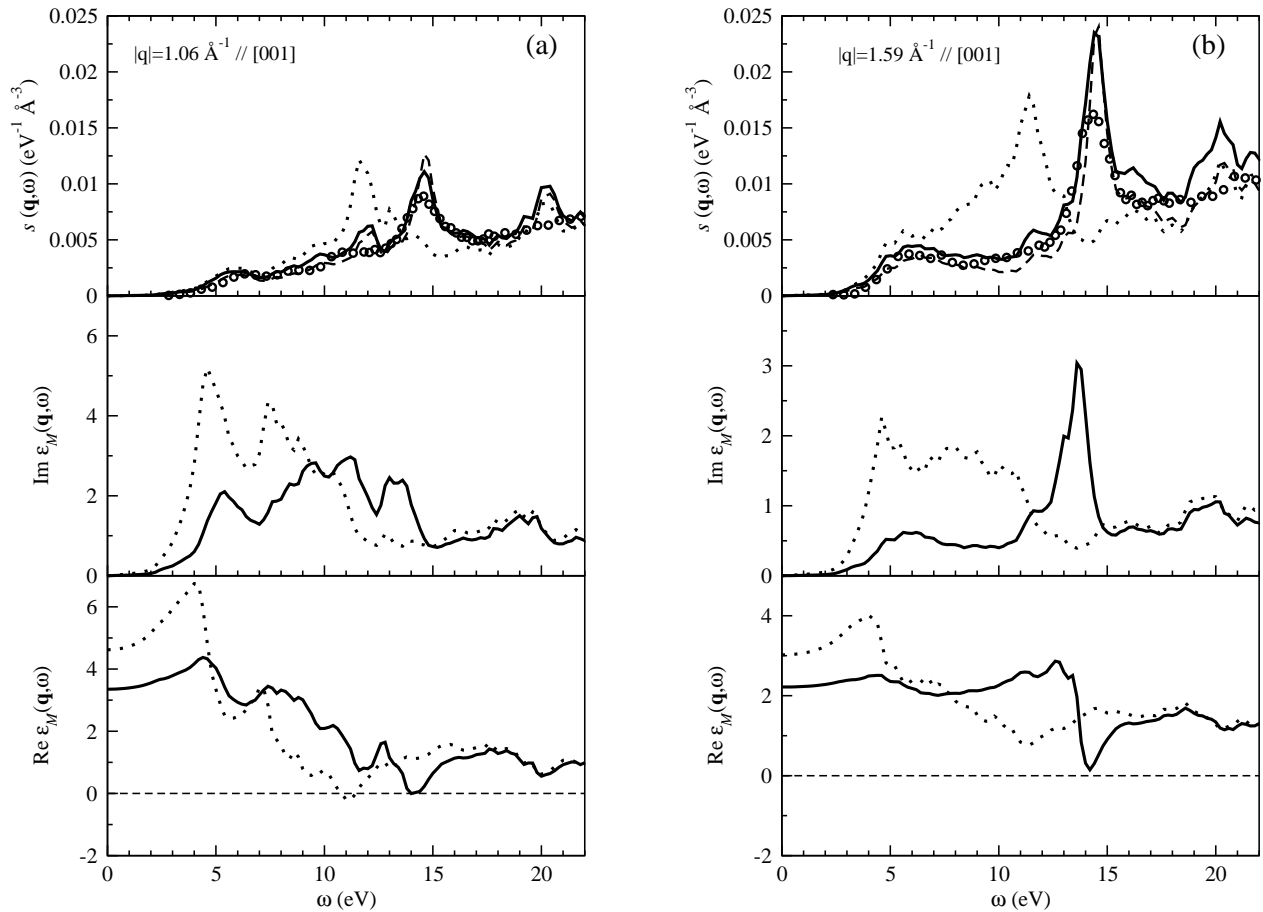}
\caption{\label{fig1} Dynamical structure factor per unit
volume $s({\bf q},\omega)$ (top panel) and macroscopic dielectric
function $\epsilon_M({\bf q},\omega)$ (middle and bottom panels) of
TiO$_2$ at (a) $|{\bf q}|$=1.06~${\rm \AA^{-1}}$ and 
(b) $|{\bf q}|$=1.59~${\rm \AA^{-1}}$
along the [001] direction. The solid (dotted) line represents the
calculated ALDA spectrum with (without) LFE. The dashed line denotes
the calculated RPA result with LFE. Open circles shows the IXS
measurements normalized to the same absolute units as the 
theoretical spectra. }
\end{figure*}

For vanishing and small momentum transfers our calculations esentially
reproduce the calculations reported in Refs.~\onlinecite{tio2mo} and
\onlinecite{tio2reining}.  Besides, for large wave vectors not
addressed before 
we find large crystal local-field effects, as shown by
the dynamical structure factor per unit volume $s({\bf
q},\omega)=S({\bf
q},\omega)/\Omega$ of rutile TiO$_2$ which we plot in Fig.~\ref{fig1}
for two values of the wave vector and for energies below
the Ti  M-edge.
IXS measurements of this quantity are
scaled to absolute units by first performing an absolute measurement
on Al
(with the use of the $f$ sum-rule)\cite{alnorm} and then scaling by
the ratio
of absorption
coefficients for Al and TiO$_2$. Considering the fact that  the
comparison
between the calculations and measurements is in absolute units, with
no
adjustable parameters, the agreement between theory and experiment is
remarkable.

The top panels of Fig.~\ref{fig1} show that in
the absence of
crystal local-field effects (dotted lines) a {\it simple} plasmon peak
is
present at 12~eV, which is well defined for wave vectors up to a
critical
value of $\approx 1.5\,{\rm\AA}^{-1}$ where the plasmon excitation
enters the
continuum of intraband particle-hole excitations. In the presence of
local-field effects (solid lines), this plasmon is still
present at $|{\bf q}|=1.06\,{\rm\AA}^{-1}$ (see Fig.~\ref{fig1}(a))
but it is
completely damped at larger wave vectors (see Fig.~\ref{fig1}(b));
in {\it addition}, a sharp loss peak emerges at 14~eV, which brings
our calculated energy-loss function into good agreement with
experiment.

We have also carried out calculations and measurements of the
dynamical structure factor for larger values of the momentum
transfer. We have found that the {\it new} peak at 14~eV, which only
appears at non-zero wave vectors, disappears at surprisingly large
wave vectors of the order of 3~${\rm\AA}^{-1}$ without significant
broadening. The oscillator strength of this peak, which is
qualitatively different from the collective excitatios in simple
materials, thus features a remarkable, non-monotonic dependence on
the wave vector. 

With the aim of establishing the nature of the large crystal
local-field
effects that yield the characteristic energy-loss function at 14 eV,
we have
considered the macroscopic dielectric function $\epsilon_M({\bf
q},\omega)$,
which within the first BZ is $\epsilon_M({\bf
q},\omega)=1/\epsilon_{{\bf 0},{\bf
0}}^{-1}({\bf q},\omega)$. Collective excitations occur at energies
where both the real
and the imaginary part of this dielectric function are close to zero.

The bottom and central panels of Fig.~\ref{fig1}
exhibit the real
and imaginary parts of $\epsilon_M({\bf q},\omega)$. At the
smallest wave vectors (not plotted here), crystal local-field effects
are
small and the real part of the macroscopic dielectric function is zero
at 12~eV where the
imaginary part is small, thereby yielding a plasmon
peak in the corresponding energy-loss function. However, as the wave
vector
increases, the rapid variation of microscopic electric fields acting
on
localized $d$ states lead to a dramatic redistribution of the strength
in the
imaginary part of the macroscopic dielectric function (see middle
panels of Fig.~\ref{fig1}). As a result, collective
excitations are considerably damped at $12\,{\rm eV}$ where Landau
damping is now efficient.
Instead, a well defined
collective oscillation occurs just above the {\it new} continuum of
particle-hole excitations. Indeed, the Kramers-Kronig relation between
the
real and imaginary parts of the macroscopic dielectric function
guarantees
that the real part has a dip whenever the imaginary part has a jump,
and the
result is a pronounced energy-loss peak (see upper panels of
Fig.~\ref{fig1})
that is responsible for the realization of a surprisingly sharp
collective
mode at $14\,{\rm eV}$. We have also carried out calculations of the
dynamical
structure factor of VO$_2$ (in the rutile structure), which
also
combines strong inhomogeneities of the Fermi sea with the presence of
localized $d$
states below and above the Fermi level, and  we have found that this
material also
exhibits a non-monotonic sharp collective mode at low energies and
large momentum
transfers.   

\begin{figure}
\includegraphics*[width=0.95\linewidth]{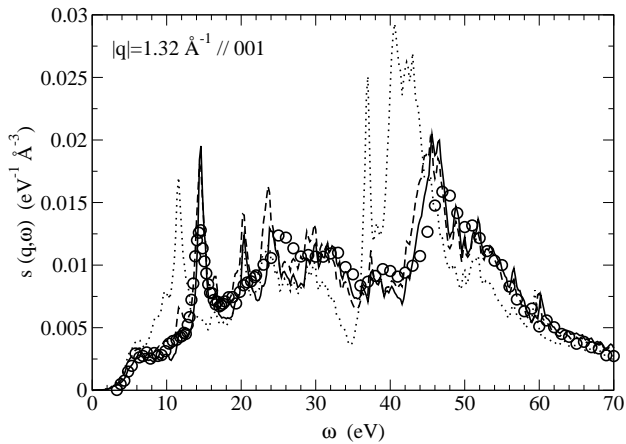}
\caption{\label{fig:sqwmed}
Dynamical structure factor of TiO$_2$ at $|{\bf
q}|=1.32$~${\rm\AA}^{-1}$ along the
[001]
direction. The solid (dotted) line represents the calculated ALDA
spectrum with
(without) crystal local-field effects. The dashed line represents the
calculated RPA spectrum with crystal local-field effects included.
Open
circles show the IXS measurements normalized to the same absolute
units as the
theoretical spectra.}
\end{figure}

We close this paper with a comparison of ALDA calculations and IXS
measurements
of the dynamical structure factor of TiO$_2$ at energies both below
and above
the Ti M-edge. Figure~\ref{fig:sqwmed} represents the results we
have
obtained for $s({\bf q},\omega)$ with $|{\bf
q}|=1.32$~${\rm\AA}^{-1}$, 
showing that when crystal local-field effects are
included the overall agreement between theory and experiment is very
good.

At low energies below the Ti M-edge, which is located at $\sim
40\,{\rm eV}$,
crystal local-field effects yield a well-defined
collective excitation at $14\,{\rm eV}$, as discussed before. At
energies
above the Ti M-edge, a prominent structure is visible, which can be
attributed
to transitions from the occupied Ti semicore $3p$ states to the lowest
conduction bands. In the absence of local-field effects
(dotted line),
the onset of these transitions would occur at $36\,{\rm eV}$. However,
local-field effects bring the onset of the M-edge to $\sim 42\,{\rm
eV}$ and reduces the corresponding peak heights, in close agreement
with
experiment. Nevertheless, there is still a small mismatch between the
converged ALDA calculation and the measured spectrum at the position
of the
main semicore peak. This mismatch stems from the fact that our
calculated LDA
ground state exhibits an upper semicore edge that is located a few eV
higher
than observed experimentally. Fig.~\ref{fig:sqwmed} also shows that
many-body XC effects not included in the RPA (dashed line) do not
affect the
low-energy collective excitation at $14\,{\rm eV}$. These effects,
however,
slightly modify the structure of the collective excitation at
$\sim$25~eV (see
also Figs.~\ref{fig1}(a) and \ref{fig1}(b)) and yield a small shift of
the semicore
excitation onset and main peak towards lower frequencies.   

In summary, we have presented a combined theoretical and experimental
investigation of the dynamical structure factor of rutile TiO$_2$ at
large momentum
transfers. We have found that large crystal local-field effects in
this wide-band-gap
semiconductor yield a  sharp loss peak at $14\,{\rm eV}$. An analysis
of the macroscopic dielectric function of this material leads us to
the
conclusion that this loss peak is originated in a collective
excitation
which remains present at surprisingly large momentum transfers and
whose spectral weight features a
remarkable non-monotonic dependence on the wave vector. This feature
is
exhibited by both TDDFT calculations and IXS measurements, which show
a
remarkable agreement when presented in absolute units with no
adjustable
parameters.

I.G.G. and J.M.P. acknowledge partial support by the Basque 
Unibertsitate, Hezkuntza  eta Ikerketa Saila, the UPV/EHU, and the
MCyT. W.K.
acknowledges support from the U.S. DOE under Contract No.
DE-AC02-98CH10886. 
A.G.E. acknowledges support from NSF ITR DMR-0219332.  
ORNL research sponsored by the DOE,
Office of Science, DMS under contract with
UT-Battelle, LLC;  the UNICAT beamline supported by the FS-MRL, ORNL,
NIST,
and UOP Res.; the Advanced Photon Source (APS) supported by the DOE.

\end{document}